\documentclass{article}

\usepackage{PRIMEarxiv}

\usepackage[utf8]{inputenc} % allow utf-8 input
\usepackage[T1]{fontenc}    % use 8-bit T1 fonts
\usepackage{hyperref}       % hyperlinks
\usepackage{url}            % simple URL typesetting
\usepackage{booktabs}       % professional-quality tables
\usepackage{amsfonts}       % blackboard math symbols
\usepackage{nicefrac}       % compact symbols for 1/2, etc.
\usepackage{subfig}
\usepackage{booktabs}
\usepackage{multicol}

\usepackage{microtype}      % microtypography
\usepackage{lipsum}
\usepackage{fancyhdr}       % header
\usepackage{graphicx}       % graphics
\graphicspath{{media/}}     % organize your images and other figures under media/ folder

\newcommand{\bb}{\begin{equation}}
\newcommand{\ee}{\end{equation}}
\newcommand{\ug}{\; = \;}

\title{
\thanks{}: 
\textbf{}}

\title{Microstructured optical beams with azimuthal polarization in stratified absorbent media}

\author{Grazielle de A. Louren\c{c}o-Vittorino and Michel Zamboni-Rached\\
University of Campinas, Campinas, S\~ao Paulo, Brazil}
\begin{document}
\maketitle

\begin{abstract}

%\begin{keyword}

We present an analytical method to achieve highly non-paraxial, azimuthally polarized structured beams that, when propagating through an absorbing stratified media, can assume in the last semi-infinite layer approximately any desired longitudinal intensity pattern within spatial regions few times larger than the wavelength. The possibility of managing the properties of a highly non-paraxial beam under adverse conditions, such as multiple reflections in stratified structures and energy loss to the material media, may be of great importance in many different optical applications, like trapping and micro-manipulation, remote sensing, thin films, medical devices, medical therapies and so on.

%We believe our results contributes for further developments in areas such as thermotherapy, medical  devices, remote sensing, military
%applications, noninvasive measurements, etc.
\end{abstract}

%\end{frontmatter}
\section{Introduction}

The possibility of getting some control on light propagation is always desirable since it expands the prospects for technological developments. In this context, a sophisticated class of localized waves (LWs) \cite{Localized-Waves08}\cite{Localized-Waves13}\cite{Zamboni-Rached2008} named Frozen Waves (FWs) \cite{Zamboni-Rached2004}\cite{Zamboni-Rached2005}  - which are composed by suitable superposition of Bessel beams - has been constantly studied and improved, providing impressive results in optics and acoustic and being experimentally verified \cite{Zamboni-Rached2006}\cite{Zamboni-Rached2009}\cite{Zamboni-Rached2010}\cite{Corato Zanarella}\cite{Vieira2012}\cite{Vieira2015}. More specifically, the FW method allows us to perform a management of the longitudinal intensity pattern of the resulting beam through a suitable superposition of co-propagating Bessel beams.   %by choosing a morphological function at will.

 In spite of being originally developed for homogeneous media, some advances have been accomplished for inhomogeneous cases \cite{Lourenco-Vittorino}, allowing to design a paraxial FW-type beam that, when propagating through a multilayered nonabsorbing media, is able to compensate the effects of inhomogeneity, providing the desired structured optical beam in the last medium.

In this paper, we extend the aforementioned methodology and propose an approximate analytical method to obtain a highly non-paraxial beam with azimuthal polarization that, at normal incidence on an absorbing stratified medium, will provide in the last semi-infinite absorbing layer a structured beam endowed with a predefined micrometer intensity pattern.

%\h In this paper, we extend the aforementioned methodology and propose an approximate analytical method to obtain a highly non-paraxial (vectorial) beam that, at normal incidence on an absorbing stratified media, will provide in the last medium a structured beam endowed with a predefined micrometer intensity pattern. It will be analyzed the cases of linear, azimuthal and radial polarizations, where the vector fields are obtained via Maxwell equations from the scalar solution of a FW beam.

%%\section{A brief discussion about a transverse wavenumber approximation }

%\section{An approximation when considering an absorbing stratified medium}

\section{A useful approximation for the transverse wave number of a Bessel beam in an absorbing medium}

First off all, before presenting our theoretical methodology, it is necessary a brief clarification about an approximation adopted for the beam solutions we are going to work with.

%For now, we will devote our attention to the transverse wavenumber $h$ and the other terms that composes the solution - phase and spectrum, in the case of the continuous superposition - we will address later on.

Let us consider a linear, homogeneous absorbing medium, with complex refractive index $n=n_r+in_i$, and a scalar wave field with azimuthal symmetry $\Psi = \textrm{exp}(-i \omega t)\psi(\rho,z)$, where $\psi$ obeys the Helmholtz equation $\partial^2\psi/\partial\rho^2 + (1/\rho)\partial\psi/\partial\rho + \partial^2\psi/\partial z^2 + n^2k_0^2\psi=0$, where $k_0 = \omega/c$ and $c$ is the light speed.

The zero-order Bessel beam solution in this case is given by

%For the sake of simplicity, let us consider a single Bessel beam in a loss medium (the time harmonic term $\textrm{exp}(- i \omega t)$ is omitted throughout the entire paper),

\bb
\psi=J_0(h\rho)\exp(i\beta z)
\ee
with
\bb
\beta^2+h^2=n^2k_0^2
\label{relacao_dispersao}
\ee
where $h$ and $\beta$ are the transverse and longitudinal wavenumbers, respectively.

%, $n=n_r+in_i$ is the complex refractive index ($r$ and $i$ refers to the real and imaginary parts) and $k_0=\omega/c$.
%\h On assuming a real transverse wavenumber, $h=an_r\omega/c$ where $0\leq a \leq 1$, and given the longitudinal wavenumber $\beta=\beta_r+i\beta_i$, from $(\ref{relacao_dispersao})$ we have that

By writing the complex longitudinal wavenumber as $\beta=\beta_r+i\beta_i$, we have from Eq.(\ref{relacao_dispersao}) that

\bb
h=\sqrt{n^2k_0^2-\beta^2}=\sqrt{(n_r^2-n_i^2)k_0^2-(\beta_r^2-\beta_i^2)+i(2n_rn_ik_0^2-2\beta_r\beta_i)}
\label{h}
\ee

Now, by demanding that $h$ is real, we must have in Eq.(\ref{h}) that $2n_rn_ik_0^2=2\beta_r\beta_i \rightarrow \beta_i=n_r n_ik_0^2/\beta_r$. If we assume $\beta_r=bn_rk_0$, with $-1\leq b\leq 1$, Eq.$(\ref{h})$ can be written as
\bb
h=\sqrt{n_r^2k_0^2-b^2n_r^2k_0^2+\left(\frac{1}{b^2}-1\right)n_i^2k_0^2}
\label{h(b)}
 \ee

The ratio between the magnitudes of the second and third term of the radicand is $<<1$ if

\bb
\frac{\left(\frac{1}{b^2}-1\right)n_i^2}{b^2n_r^2}<<1 \,\,\, \Rightarrow \frac{b^2}{\sqrt{1-b^2}}>>\frac{n_i}{n_r} \,\,\, \Rightarrow \frac{\beta_r^2}{\sqrt{n_r^2k_0^2-\beta_r^2}}>>n_ik_0
\label{condicao}
\ee
and, in such cases, we can write

%and, since $b=\beta_r/(n_rk_0)$, inequation (\ref{comparacao}) became
%\bb
%\frac{\beta_r^2}{\sqrt{n_r^2k_0^2-\beta_r^2}}>>n_ik_0
%\label{condicao}
%\ee
%
%\h Thus, the condition $(\ref{condicao})$ must be satisfied in order to obtain

\bb
h\approx\sqrt{n_r^2k_0^2-b^2n_r^2k_0^2}=\sqrt{n_r^2k_0^2-\beta_r^2}
\label{h_aproximado}
\ee

%\h When we consider a stratified media, we have to bear in mind that the transverse wavenumber is a conserved quantity in all $M$ layers due to the boundary conditions, and ($\ref{h_aproximado}$) will be a good approximation if
%$(\ref{condicao})$ is observed in all layers. We will certify our approach by checking this condition, for that, we first get all of $\beta_{rm}$ through $(\ref{relacao_dispersao})$, that is, $\beta_{rm}=Re(\sqrt{n_m^2\omega^2/c^2-h^2})$ where $m=1 ,2,...M$. With this result at hand, we can verify for those layers with loss if

The transverse wave numbers of the Bessel beams considered in this work will be particularly important because, as we are going to consider normal incidences with the interfaces of the stratified medium, they are conserved and, thus, will be fundamental in obtaining the longitudinal wave numbers of the Bessel beams in any layer of that medium.

 In this context, the approximation (\ref{h_aproximado}) is very useful because, as we will see, it makes possible the structured optical beams obtained through our method be described by analytical solutions of Maxwell's equations, without the need of numerical solutions/simulations. Due to this, in any layer of the stratified medium we are dealing with, we will certify that\footnote{Such a condition is satisfied in most cases where $n_i<<n_r$.}

\bb
\frac{\beta_{rm}^2}{\sqrt{n_{rm}^2k_0^2-\beta_{rm}^2}}>>n_{im}k_0
\label{condicao1}
\ee

where $\beta_{rm}=Re(\sqrt{n_m^2\omega^2/c^2-h^2})$ is the longitudinal wave number of the Bessel beam in the mth layer, whose refractive index, $n_m$, possesses real and imaginary parts given by $n_{rm}$ and $n_{im}$, respectively.

%%where $n_{im}$ is the imaginary part of the complex refractive index of the media $m=1 ,2,...M$.

%\h In summary, ($\ref{condicao1}$) occurs for media whose imaginary part of the refractive index has small value as adopted in this work, where $n_i$ is about $10^3$ times smaller than $n_r$. Therefore, within the examples to be discussed in this article the $h$ given by $(\ref{h_aproximado})$ is a good approximation.

\section{The method}

%\subsection{\textbf{Characterization of the desired beam}}

%\h The aim of this work is, given a absorbing stratified medium with M layers (the first and the last ones being semi-infinite), to obtain an incident optical beam in such a way that in the last medium we have a non-paraxial, azimuthally polarized and microstructured beam, capable of assuming a longitudinal intensity pattern chosen on demand.

Here, we present the method by considering a scalar field which, in the next section, will be used in obtaining the desired results for an azimuthally polarized beam.
%
%
%
%Here, we will present the method
%
%In the next subsection we will consider the scalar case, whose results will then be used in the development of the vectorial one.
%
%\subsection{\textbf{Characterizatio}}
%
%\h Due to the non-paraxial character of the problem, the vectorial nature of the beam has to be taken into account, something that will be done in subection $XX$, where the desired structured beam will be obtained as a (nearly exact) solution of Maxwell's equations. For now, we will consider the scalar case, whose results will then be used in the development of the vectorial case.

Let us consider an absorbing stratified medium with M layers, the first and the last ones being semi-infinite. We wish to construct an incident scalar wave field in such a way that in the last medium we have a non-paraxial microstructured scalar beam, capable of assuming a longitudinal intensity pattern chosen on demand.

%\h In order to obtain the necessary scalar incident beam, we first need to describe the characteristics of that transmitted ($\psi^T$, the desired beam) to the last medium of the stratified structure. To simplify the notation, we will adopt the subscripts $1$ in the longitudinal wavenumber, $\beta$, k and $n$  when we refer to the first medium, and the subscripts $r$ and $i$ in reference to the real and imaginary parts, respectively, of the complex $\beta$, k and $n$ in the last medium. In addition, let us assume that all layers, besides being linear and isotropic, can be absorbing (except the first medium which we assume to be lossless), being the complex refractive index of the last medium given by $n_M=n_r+in_i$.

 In order to get this result, we first need to describe the characteristics of the desired beam, $\psi^T$, transmitted to the last medium of the stratified structure. To simplify the notation, we will adopt the subscripts $1$ for the longitudinal wavenumber $\beta_1$, $k_1$ and $n_1$ for the longitudinal wavenumber, wavenumber and refractive index in the first medium, respectively, while to the last medium we will use just $\beta$, $k$ and $n$, with the subscripts $r$ and $i$ in reference to the real and imaginary parts when necessary.

%when referring to the first medium, and the subscripts $r$ and $i$ in reference to the real and imaginary parts, respectively, of the complex $\beta$, $k$ and $n$ in the last medium. In addition, let us assume that all layers, besides being linear and isotropic, can be absorbing (except the first medium which we assume to be lossless), being the complex refractive index of the last medium given by $n_M=n_r+in_i$.

Let us consider as an approximated\footnote{Here we will use the approximate form of the transverse wavenumber, given by eq.(\ref{h_aproximado})} solution to the scalar wave equation a continuous superposition of zero-order Bessel beams over the longitudinal wavenumber $\beta_r$:

 \bb
 \begin{array}{clr} 
 \psi^T(\rho,z) & =\int_{-k_r}^{k_r}S(\beta_r)J_0\left(\rho\sqrt{(n_r^2-n_i^2)k_0^2-\left[\beta_r^2-\left(\frac{n_r n_i k_0^2}{\beta_r}\right)^2\right]}\,\,\right)
 \exp(i\beta_r z)\exp\left(- \frac{n_rn_ik_0^2}{\beta_r} z\right) \textrm{d}\beta_r \\
 \\
 & \approx \, \exp(-\bar{\beta}_i z)\int_{-k_r}^{k_r}S(\beta_r)J_0\left(\rho\sqrt{k_r^2-\beta_r^2}\,\right)
 \exp(i\beta_rz)\,\textrm{d}\beta_r \,\, ,  
 \label{psi_t}
 \end{array} 
 \ee
 
 being $k_0 = \omega/c$, $k_r=n_r k_0$, $\beta_r$ is the real part of the complex longitudinal wavenumber of the medium $M$, i.e, $\beta=\beta_r+i\beta_i$, whose imaginary part $\beta_{i}=k_0^2 n_{r}n_{i}/\beta_{r}$ is responsible for an exponential decay along the propagation direction. The approximation on the integral solution (\ref{psi_t}) was made by considering the approximate form of the transverse wavenumber, given by eq.(\ref{h_aproximado}), and also by considering $\beta_{i} \approx \bar{\beta}_{i}=k_0^2\frac{n_{r}n_{i}}{Q}$, where the parameter $Q=an_r k_0$ $(0<a<1)$, as we will see later, corresponds to the value of $\beta_r$ where the center of the spectrum $S(\beta_r)$ is localized.

Let us also consider the spectrum as given by the following Fourier series

\bb S(\beta_r)=\sum_{m=-\infty}^{\infty} A_m \exp\left(\frac{i2m\pi}{K_r}\beta_r\right)
\label{sbr}
\ee
with
\bb A_m= \frac{1}{K_r}F\left(-\frac{2m\pi}{K_r}\right)\label{am}\,\, , \,\,\, K_r=2k_r
\label{am}
\ee
and where $F(\cdot)$ is a function of our choice, as explained below.

 Due to Eqs. $(\ref{sbr})$ and ($\ref{am}$), it is possible to show \cite{Zamboni-Rached2008,Zamboni-Rached17} that $|\psi^T(\rho=0,z)|^2\approx \exp(-2\bar{\beta}_{i})|F(z)|^2$. Now, we choose $F(z)$, called morphological function, as given by

\bb
F(z)=f(z)\exp({\bar{\beta}}_i z)\exp(iQz)
\label{morfologica}
\ee
where the first function, $f(z)$, provides the desired beam intensity profile, the second one, $\exp({\bar{\beta}}_i z)$, assigns attenuation resistance to the beam and the third one, $\exp(iQz)$, is responsible for shifting the spectrum, ensuring that $S(\beta_r)$  gets negligible values for $\beta_r<0$, ensuring only forward propagating Bessel beams in superposition ($\ref{psi_t}$). Actually, we will impose an even more restrictive condition over $S(\beta_r)$, that is

\bb S(\beta_r)\approx0 \,\,\, \textrm{for}\,\,\, \beta_r<\sqrt{k_r^2-k_1^2} \,, \label{cond_betar}\ee

because $\beta_1=\sqrt{(k_1^2-k_r^2)+\beta_r^2}$ (the longitudinal wavenumber in the first medium) is imaginary for $\beta_r<\sqrt{k_r^2-k_1^2}$.

%In short, we assume for $S(\beta_r)$ that:
%\begin{enumerate}
%\item $S(\beta_r)\approx0$ for $\beta_r<0$;
%\item $S(\beta_r)\approx0$ for values of $\beta_r<\sqrt{k_r^2-k_1^2}$, because for those values of $\beta_r$ , $\beta_1$ (longitudinal wavenumber in medium $1$) is imaginary, given that $\beta_1=\sqrt{(k_1^2-k_r^2)+\beta_r^2}$.
%\end{enumerate}

Finally, it is possible to show \cite{Zamboni-Rached2008,Zamboni-Rached17} that the integral solution $(\ref{psi_t})$ results to be a discrete superposition of Mackinnon-type beams:
\bb
\psi^T(\rho,z)\approx\ \exp(-\bar{\beta}_i z)\sum_{m=-\infty}^{\infty}A_m sinc\left[k_r^2\rho^2+\left(k_rz+m\pi\right)^2\right] \,,
\label{psi_t_mack}
\ee
where $sinc(\cdot)$ is the sinc function.   %% \textrm{sinc}

\

\emph{Calculating the incident beam:}

\

With the scalar FW beam characterized in the last medium as we wish, Eq.(\ref{psi_t_mack}), we can estimate the necessary incident beam $\psi^I$ in the first (nonabsorbing) medium as

\bb
\psi^I(\rho,z)\approx\ \int_{0}^{k_r}\frac{S(\beta_r)}{\tau(\beta_1)}J_0(\rho\sqrt{k_r^2-\beta_r^2}\,)
e^{i\beta_1}d\beta_r \,
\label{psi_i}
\ee
where the square root in the argument of the Bessel function is the transverse wavenumber, which is conserved due to the boundary conditions, and $\tau(\beta_1)$ is the effective transmission coefficient, referring to the entire stratified structure, given as a function of $\beta_1$. A description of how to obtain such transmission coefficient through a transfer-matrix method can be found in the appendix.

Since $\beta_1=\sqrt{(k_1^2-k_r^2)+\beta_r^2}$, there are two possible situations in the superposition given by Eq.$(\ref{psi_i})$: if $n_1>n_r$, $\beta_1$ is always real, but if $n_1<n_r$ then:

\begin{enumerate}
	\item $\beta_1=i\sqrt{(k_r^2-k_2^2)-\beta_r^2}$~~~if~~ $\beta_r<\sqrt{k_r^2-k_1^2}$ or
	\item $\beta_1=\sqrt{(k_1^2-k_r^2)+\beta_r^2}$~~~~if~~ $\beta_r>\sqrt{k_r^2-k_1^2}$
\end{enumerate}

Here, we restrict to the case where $n_1=1$ (vacuum), so the incident beam will be:
\bb
\psi^I(\rho,z)\approx\ \int_{0}^{\sqrt{k_r^2-k_1^2}}\frac{S(\beta_r)}{\tau(\beta_1)}J_0(\rho\sqrt{k_r^2-\beta_r^2}\,)
e^{-\sqrt{(k_r^2-k_1^2)-\beta_r^2}z}d\beta_r
\\
+\int_{\sqrt{k_r^2-k_1^2}}^{k_r}\frac{S(\beta_r)}{\tau(\beta_1)}J_0(\rho\sqrt{k_r^2-\beta_r^2}\,)
e^{i\sqrt{(k_1^2-k_r^2)+\beta_r^2}z}d\beta_r
\label{psi_ifull}
\ee
Due to condition Eq.(\ref{cond_betar}), the first integral in Eq.$(\ref{psi_ifull})$ is negligible compared to the second one. So, by considering just the second integral in eq.(\ref{psi_ifull}) and changing the integration variable from $\beta_r$ to $\beta_1$ by using that $\beta_r=\sqrt{(k_r^2-k_1^2)+\beta_1^2}$, we can write

\bb
\psi^I(\rho,z)\approx\int_{0}^{k_1}S'(\beta_1)J_0(\rho\sqrt{k_1^2-\beta_1^2}\,)
\exp(i\beta_1 z)d\beta_1
\ee
where
\bb
S'(\beta_1)= \frac{S[\beta_r(\beta_1)]}{\tau(\beta_1)}\frac{\beta_1}{\sqrt{(k_r^2-k_1^2)+\beta_1^2}}
\label{s(b1)}
\ee
and
\bb S[\beta_r(\beta_1)]=\sum_{m=-\infty}^{\infty} A_m \exp\left[\frac{i2m\pi}{K_r}\beta_r(\beta_1)\right]
\label{coef_s(b1)}\ee

 By using the Heaviside function,

\bb
H(\beta_1)=\left\{
\begin{array}{c}
	\ 1~~~~~~~~if~~~\beta_1>0\\
	\ 0~~~~~~~~if~~~\beta_1<0 
\end{array}\right.
 \label{quadrado}
\ee

we can write
 \bb
\psi^I(\rho,z)\approx\ \int_{-k_1}^{k_1}H(\beta_1)S'(\beta_1)J_0(\rho\sqrt{k_1^2-\beta_1^2}\,)
\exp(i\beta_1z)d\beta_1
\label{psi_i_integral}
\ee

whose solution is

\bb
\psi^{I}(\rho,z)\approx\ 2k_1\sum_{g=-\infty}^{\infty}D_g sinc\sqrt{k_1^2\rho^2+\left(k_1z+g\pi\right)^2} \,,
\label{psi_i_mack}
\ee

where we have used that

\bb
H(\beta_1)S'(\beta_{1})=\sum_{g=-\infty}^{\infty}D_g \exp\left(\frac{i2g\pi }{K_1}\beta_1\right) \,,
\label{sb1}
\ee

with

\bb
D_g=\frac{1}{2k_1}\int_{-k_1}^{k_1}H(\beta_1)S'(\beta_{1})\exp\left(-\frac{i2g\pi }{K_1}\beta_1\right)d\beta_1
\label{dg}
\ee

In summary, in order to have the desired wave field in the last medium, i.e., a scalar beam microstructured according to a morphological function $F(z)$ and whose analytic solution is given by Eq.(\ref{psi_t_mack}), the incident wave at the first interface of the stratified structure has be given by the analytic solution Eq.(\ref{psi_i_mack}).

\section{The method applied to a non-paraxial and azimuthally polarized optical beam}

When dealing with vector beams it is always mandatory to keep in mind that the behavior of TE and TM beams are different with respect to the phenomena of reflection and refraction.
In the case of a TE Bessel beam at normal incidence on a plane interface separating two dielectrics, the reflection and transmission coefficients are equal to those of a scalar Bessel beam (of the same cone angle) also at normal incidence. This fact will be used here.
%\h Because we are dealing with a highly non-paraxial beams, we must consider the vectorial nature of the field.

The aim of this work is, given a absorbing stratified medium with M layers (the first and the last ones being semi-infinite), to obtain an incident optical beam in such a way that in the last medium we have a non-paraxial, azimuthally polarized and microstructured beam, capable of assuming a longitudinal intensity pattern chosen on demand.

In this section, we will obtain such microstructured electromagnetic beam via Maxwell's equations, taking as a starting point the scalar solution presented in the previous section.

%More specifically, it will be achieved the analytical solution of the incident field which yield the desired transmitted one. , considering linear, azimuthal and radial polarizations. The corresponding magnetic field can be calculated from Faraday’s law.  $\textbf{B}(\textbf{r})=-\frac{i}{\omega}\nabla \times \textbf{E}(\textbf{r})$.

Although we do not present the calculation of the reflected beam (since we are interested in the transmitted wave), it will be shown in the figures since we think that such information contributes to a better understanding of the phenomena studied here.

Let us consider a stratified medium formed by $M$ layers with refractive indices $n_m$ $(m=1,2,…,M)$ and with their interfaces located at the positions $z=d_1,d_2,…,d_{M-1}$.
See Fig. 1.

%Here we adopt $\lambda=532\mu$m (in vacuum).

\begin{figure}[!htb]
	\centering
	\subfloat{
		\includegraphics[height=3cm]{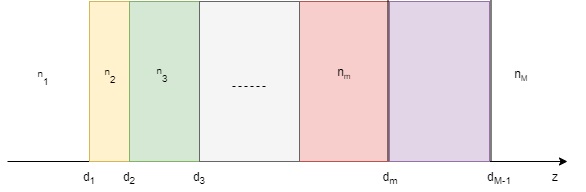}
	}
	
	\caption{Schematic representation of the stratified media}
	\label{meio_estrat}
\end{figure}

%************************************************************************

A optical beam with azimuthal polarization and azimuthal symmetry, $\textbf{E}=E_{\phi}(\rho,z)\textrm{exp}(-i \omega t)\hat{\phi}$, must obey

\bb \frac{\partial^2 E_{\phi}}{\partial \rho^2} + \frac{1}{\rho}\frac{\partial E_{\phi}}{\partial \rho} - \frac{E_{\phi}}{\rho^2} + \frac{\partial^2 E_{\phi}}{\partial z^2} + n^2k_0^2E_{\phi}=0 \,\, , \label{eqephi}\ee
whose the simplest solution is a first order Bessel-type-beam, i.e, $E_{\phi}(\rho,z)=J_1(h\rho)\exp(i\beta z)\hat{\phi}$.

Here, we could follow a procedure similar to that of the scalar case presented in Section $3$ and write the transmitted beam (i.e., the beam in the last medium) as $\textbf{E}^T=E_{\phi}^T(\rho,z)\textrm{exp}(-i \omega t)\hat{\phi}$ being $E_{\phi}^T(\rho,z)$ a superposition similar to Eq.(\ref{psi_t}), only replacing the zero-order Bessel function $J_0(\cdot)$ by the first-order one $J_1(\cdot)$. It turns out, however, that in this case there is no known analytic solution for that integral when $S(\beta_r)$ is given by Eq.(\ref{sbr}).

To overcome this issue, we will adopt the following strategy: it is very simple to verify that differentiating the solution of the Helmholtz equation $\psi =J_0(h \rho) \exp(i\beta z)$ (where $h = \sqrt{k^2 - \beta^2}$) with respect to $\rho$, we obtain $\partial \psi / \partial \rho = - h J_1(h \rho) \exp(i\beta z) $, which in turn is a solution of Eq. (\ref{eqephi}). Thus, it is clear that if we differentiate the integral solution (\ref{psi_t}) of the Helmholtz equation with respect to $\rho$, we will get an integral solution to the differential equation (\ref{eqephi}).

That being said, we will write $E_{\phi}^T = \xi \partial \psi^T / \partial \rho$, with $\psi^T$ given by Eq.(\ref{psi_t}) and $\zeta$ a normalization constant:

\bb
 E_{\phi}^T(\rho,z) \approx\ \xi\exp(-\bar{\beta}_i z)\int_{-k_r}^{k_r}S''(\beta_r)J_1(\rho\sqrt{k_r^2-\beta_r^2})
 \exp(i\beta_rz)d\beta_r \,\, ,
 \label{EphiT int}
\ee

with

\bb S''(\beta_r) = - \sqrt{k_r^2-\beta_r^2}\,S(\beta_r) \,\, , \label{S''}\ee
and $S(\beta_r)$ given by Eqs.(\ref{sbr},\ref{am}).

 Naturally, from Eq.(\ref{psi_t_mack}), the analytical solution to the integral solution Eq.(\ref{EphiT int}) will be given by:

\bb
E_\phi^{T}(\rho,z)\approx \xi e^{-\bar{\beta}_i z}  \sum_{m=-\infty}^{\infty} A_m \, \frac{\partial}{\partial \rho}sinc\sqrt{k_r^2\rho^2+\left(k_r z+ \pi m\right)^2} \,\,, \label{EphiT}
\ee
always keeping in mind that the coefficients $A_m$ are given by Eqs.(\ref{am},\ref{morfologica}).

The central concern now is whether the integral solution (\ref{EphiT int}) of the beam transmitted to the last semi-infinite layer enables it to have its longitudinal intensity pattern modeled at will within a micrometer spatial region. This concern occurs due to the fact that the spectrum $S''(k_z)$ in the superposition (\ref{EphiT int}) differs from the spectrum $S(k_z)$, Eqs.(\ref{sbr},\ref{am}), which enables the spatial modeling of the scalar field $\psi_T$.

 Fortunately, the factor $\sqrt{k_r^2-\beta_r^2}$ in Eq.(\ref{S''}) is not able, in general, to substantially modify the shape of $S''(k_z)$ when compared $S(k_z)$; actually, for the vast majority of cases of interest, both spectra are very similar, except for a difference in amplitude. This causes the longitudinal field pattern of $E_\phi^{T}(\rho,z)$ to be dictated by the morphological function $F(z)$, as intended. In addition, the field is no longer concentrated over the axis $\rho=0$, but it is now concentrated over a cylindrical surface of radius $\rho_{1} \approx 1.84 / \sqrt{k_r^2 - Q^2}$, where the number $1.84$ is the value of the argument that maximizes the Bessel function $J_1(.)$.

%where $\eta_{1}$ is such that $J_{1}(\eta)$ has maximum value at $\eta = \eta_{1}$.

 In this way, we can say that solution (\ref{EphiT}) represents a microstructured beam in the semi-infinite layer M, behaving according to the morphological function $F(z)$ given by Eq.(\ref{morfologica}), more specifically $E_\phi^{T}(\rho=\rho_1,z)\approx f(z)\exp(i Q z)$ with $f(z)$ and $Q$, chosen at will.

 Now, we proceed to calculate $E_\phi^{I}$, the beam incident on the first interface of the stratified medium that, ultimately, gives rise to the transmitted field $E_\phi^{T}$.

 Since the transmission and reflection coefficients of a TE Bessel-type beam at normal incidence on a plane interface are the same as those for a scalar Bessel beam, we can use the results of section $3$ and write for the incident beam:

\bb
E_\phi^{I}(\rho,z)\approx \xi  \sum_{g=-\infty}^{\infty}D'_g\frac{\partial}{\partial \rho}sinc\sqrt{k_1^2\rho^2+\left[k_1 z+\pi g\right]^2} \,\,, \label{EphiI}
\ee
where the coefficients $D'_g$ are given by

\bb
D'_g=\frac{1}{2k_1}\int_{-k_1}^{k_1}H(\beta_1)S'''(\beta_{1})\exp\left(-\frac{i2g\pi }{K_1}\beta_1\right)d\beta_1 \,\,,
\label{dg'}
\ee
where

\bb
S'''(\beta_1)= \frac{S''[\beta_r(\beta_1)]}{\tau(\beta_1)}\frac{\beta_1}{\sqrt{(k_r^2-k_1^2)+\beta_1^2}} \,\,,
\label{s'''}
\ee
with
\bb S''[\beta_r(\beta_1)]= - \sqrt{k_r^2-\beta_r^2(\beta_1)}\,\sum_{m=-\infty}^{\infty} A_m \exp\left[\frac{i2m\pi}{K_r}\beta_r(\beta_1)\right]
\label{s''(b1)}\ee
and $\beta_r(\beta_1)=\sqrt{(k_r^2-k_1^2)+\beta_1^2}$. Notice that $\tau(\beta_1)$ is the effective transmission coefficient (of the stratified medium) given as a function of $\beta_1$; it can be achieved through a transfer-matrix method, as described in the appendix.

 Using a condensed notation, we can write the incident/transmitted beam pair as:

\bb
E_\phi^{I\choose T}(\rho,z)\approx \xi{1\choose e^{-\bar{\beta}_i z}}  \sum_{{g\choose m}=-\infty}^{\infty}{D'_g\choose A_m}\frac{\partial}{\partial \rho}sinc\sqrt{{k_1^2\choose k_r^2}\rho^2+\left[{k_1\choose k_r}z+\pi {g\choose m}\right]^2}
\ee

 The magnetic field $\mathbf{B} = B_{\rho}\hat{\rho} + B_{z}\hat{z}$, obtained from the Faraday law, can be writen in the same notation as:

\bb
B_\rho^{I\choose T}(\rho,z)\approx \xi\,\frac{i}{\omega}{1\choose e^{-\bar{\beta}_i z}}  \sum_{{g\choose m}=-\infty}^{\infty}{D'_g\choose A_m}\frac{\partial^2}{\partial z\partial \rho}sinc\sqrt{{k_1^2\choose k_r^2}\rho^2+\left[{k_1\choose k_r}z+\pi {g\choose m}\right]^2}
\ee
and
\bb
B_z^{I\choose T}(\rho,z)\approx - \xi \,\frac{i}{\omega}{1\choose e^{-\bar{\beta}_i z}}  \sum_{{g\choose m}=-\infty}^{\infty}{D'_g\choose A_m}\frac{1}{\rho}\frac{\partial}{\partial \rho}\left[\rho\frac{\partial}{\partial \rho}sinc\sqrt{{k_1^2\choose k_r^2}\rho^2+\left[{k_1\choose k_r}z+\pi {g\choose m}\right]^2} \right]
\ee

In summary, for the azimuthally polarized beam, Eq.(\ref{EphiT}), microstructured according to the morphological function $F(z)$, be the one transmitted to the last medium of the stratified structure, it is required that in the first medium the incident beam be given by Eq.(\ref{EphiI}).

%% PAREI AQUI

\emph{An example:}

 Here we adopt $\lambda=532$nm (in vacuum).

Let us consider a simple stratified media composed by three layers, whose refractive index as well as the interfaces locations are depicted in Table $(1)$.

\begin{table}
	\centering
	\caption[tab2]{}
	\begin{tabular}{|l|c|c|c|}
		%\toprule[1.5pt]
		Layer $(m)$ &  Refractive index ($n_m$)  & Thickness ($\mu$m)& Interface at $z$ ($\mu$m) \\
		\midrule
		1   &  1   &        semi-infinite & $d_1=0$\\
		2   &  1.3+0.32e-3$i$ 	&   20&   $d_2=20$\\
		3   &  1.5+3e-3$i$	&   semi-infinite& -
		%\bottomrule[1.5pt]
		%\midrule
	\end{tabular}
	%	%\caption[Filósofos]{Tabela de filósofos}
	\label{tabela2}
\end{table}

%\h The field-intensity chosen for the last medium is given by two super-Gaussians as

 In this example, the chosen morphological function is
% chosen to longitudinal pattern is built over micrometer scale modulated in the last medium by

\bb
F(z)=\exp\left[-\left(\frac{z-z_0}{Z}\right)^8\right]\cos\left(\frac{2\pi z}{\Lambda}\right)\exp\left(\bar{\beta}_iz\right)\exp(iQz)
\label{f2}
\ee
with $z_0=56\mu$m, $Z=25\mu$m, $\Lambda = (5/6)Z$ and $Q = 0.97 k_r$. % $Q=0.97 n_r \omega/c$.

 The morphological function given by Eq.(\ref{f2}) means that we wish the azimuthally polarized beam transmitted to the third medium to possess a transverse raius (hollow beam) given approximately by $\rho_{1} \approx 1.84 / \sqrt{k_r^2 - Q^2} \approx 0.43\mu$m and a longitudinal intensity pattern given by an 8th-order supergaussian centered at $z=z_0=56\mu$m, with width approximately $2Z/(2)^{1/8} \approx 46\mu$m, modulated by a squared cosine function of spatial period given by $\Lambda / 2 \approx 10.4\mu$m.

%% PAREI AQUI !!! !

 Having $F(z)$ in hand, the solution for the beam in the last medium (i.e., the transmitted beam $E_{\phi}^T$) is given by Eq.(\ref{EphiT}), with the coefficients $A_m$ given by Eqs.(\ref{am},\ref{morfologica}). The value of the normalization constant $\xi$ is chosen such that the maximum intensity of the transmitted beam (in arbitrary units) is unitary.

 Figure (2a) shows the intensity of the beam transmitted to the last medium, evidencing that the field is microstructured according to the desired shape. Figure (2b) reinforces this fact by comparing the longitudinal field intensity over the cylindrical surface of radius $\rho_{1} \approx 0.43\mu$m (red line) with the intensity demanded by the morphological function (black line).

 It is interesting to note that the transmitted beam is not only resistant to the effects of diffraction, but is also resistant to attenuation (a consequence of the term $\exp\left(\bar{\beta}_iz\right)$ in the morphological function given by Eq.(\ref{f2})). Due to the absorption presented by the last medium, an ordinary optical beam would have a penetration depth given by $\delta=c/(2n_i\omega)\approx 14\mu$m, while the structured beam $E_{\phi}^T$ is able to propagate a distance $3.3$ times greater without suffering the effects of attenuation.

 We now proceed to the incident beam $E_{\phi}^I$), given by the solution (\ref{EphiI}), where the coefficients $D'_g$ are numerically calculated from Eq.(\ref{dg'}). As already stated, this is the beam that must be generated in medium 1 so that the beam transmitted to the last medium is given by $E_{\phi}^T$).

 Figure (2c) shows the intensity of the beam incident on the first plane interface and Fig(2d) shows, in logarithmic scale, the ratio $|E_{\phi}^I|^2 / |E_{\phi}^T|^2_{max}$, where $|E_{\phi}^T|^2_{max}$ is the maximum value of the transmitted beam.

 Although we have not provided the equations for calculating the field reflected by the first interface, Fig.(2e) shows the reflected beam intensity $|E_{\phi}^R|^2$, and Fig.(2f) shows, in logarithmic scale, the ratio $|E_{\phi}^R|^2 / |E_{\phi}^T|^2_{max}$.

\begin{figure}[h!]
	\centering{
		\subfloat[]{
			\includegraphics[width=0.4\textwidth]{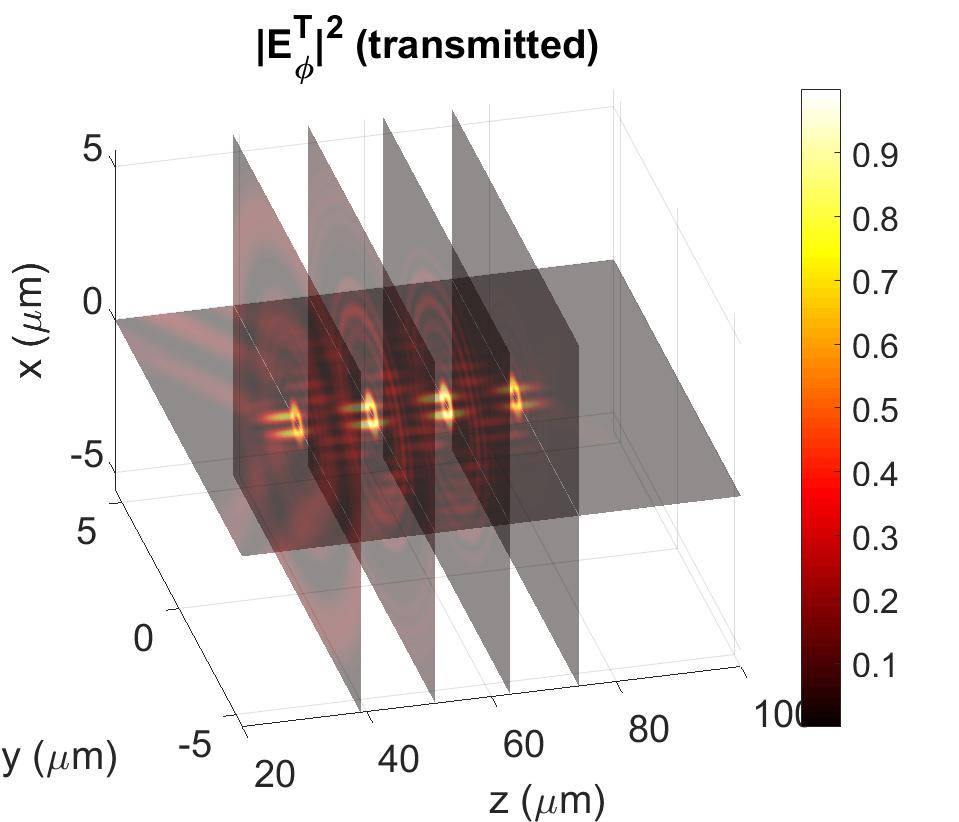}}
				\subfloat[]{
			\includegraphics[width=0.4\textwidth]{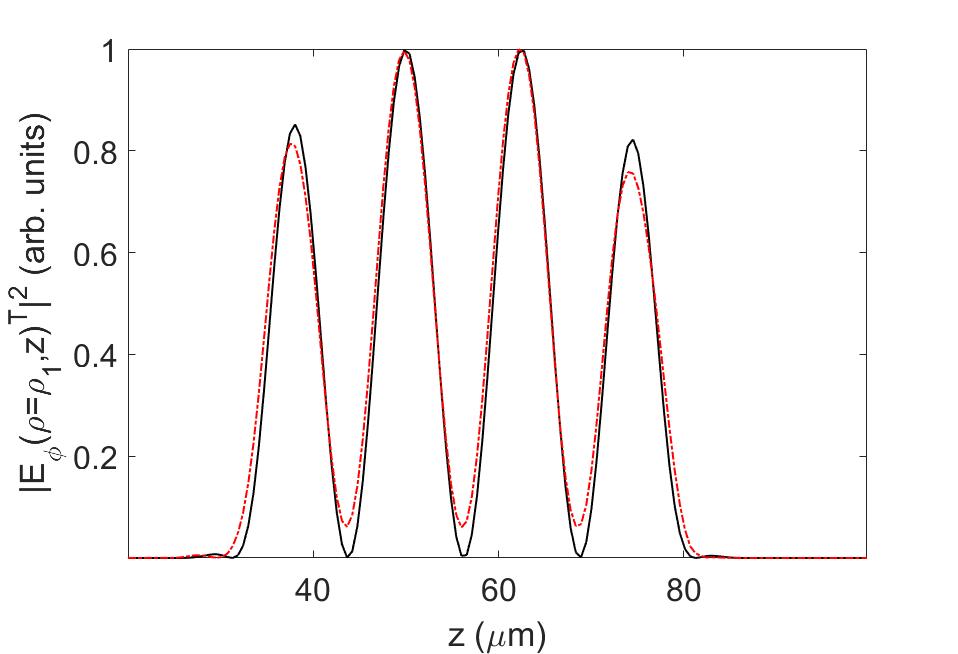}}
		\\
		\centering{
			\subfloat[]{
				\includegraphics[width=0.4\textwidth]{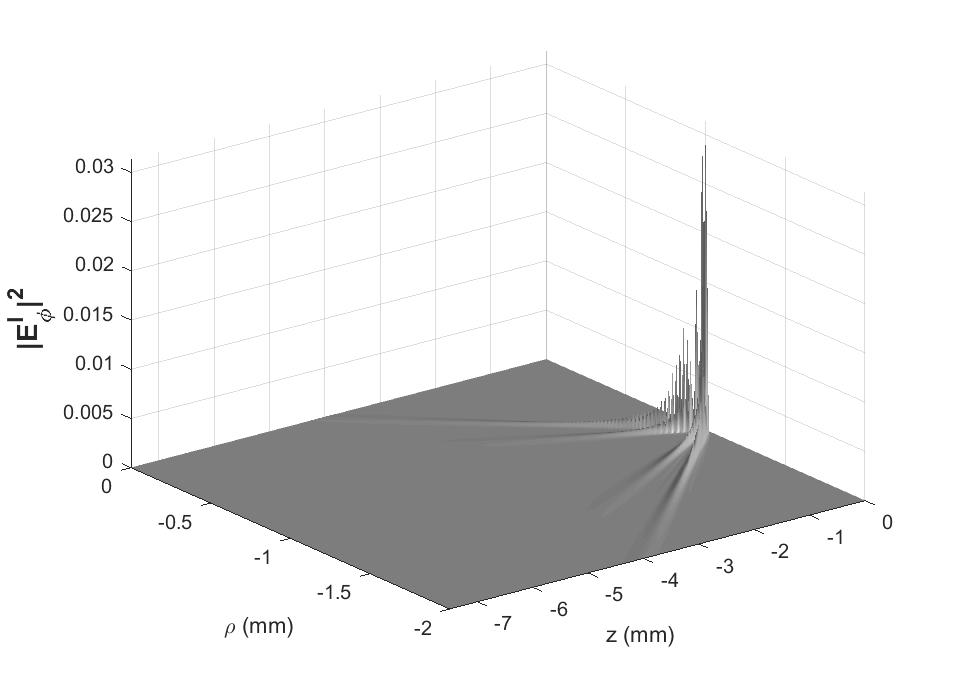}
			} %linear_2d_transmitido
			\subfloat[]{
				\includegraphics[width=0.4\textwidth]{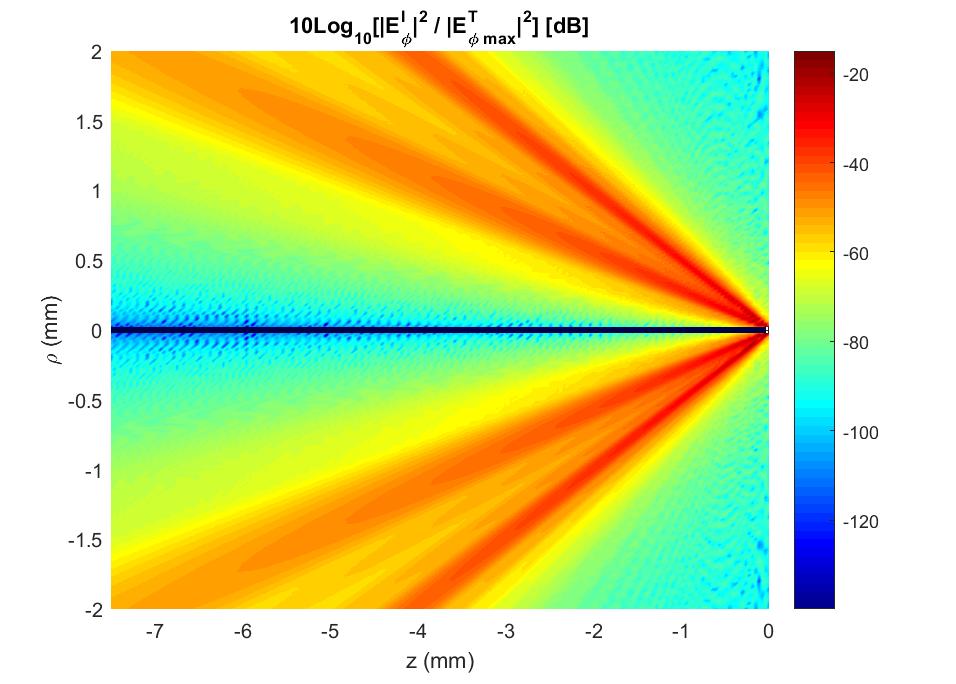}
		}}
			\\
		\centering{
			\subfloat[]{
				\includegraphics[width=0.4\textwidth]{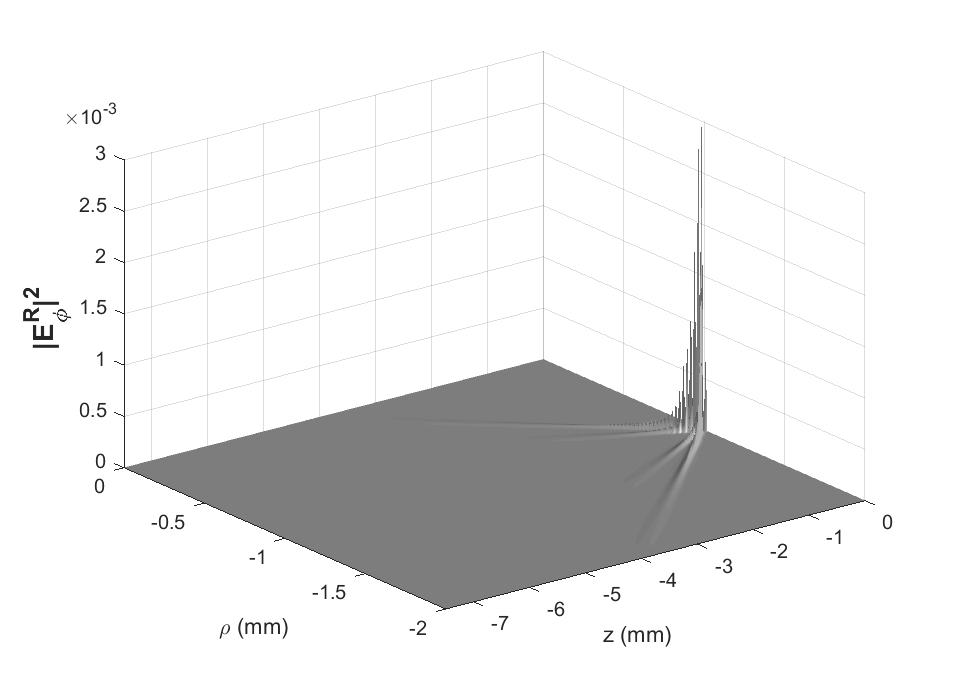}
			} %linear_2d_transmitido
			\subfloat[]{
				\includegraphics[width=0.4\textwidth]{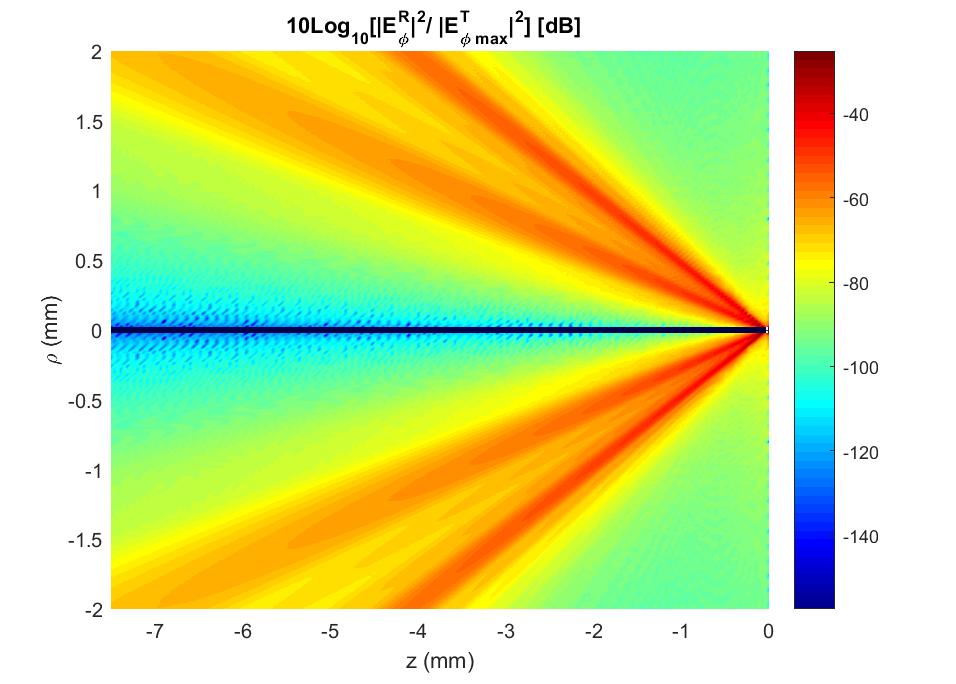}
		}}
		\caption{{ (a) The 3D intensity of the compensated transmitted azimuthal component $E_{\phi}$; (b) comparison between the on-axis longitudinal intensity $E_{\phi}$ (dotted line) and the desired pattern $|F(z)|^2$ (solid line); (c) compensated incident component $E_{\phi}$ and (d) this component in logaritmic scale; (e) reflected component $E_{\phi}$ in medium 1 and (f) this component in logaritmic scale.}
	}}
	\end{figure}

%\h Here, the FW propagated approximately $3.5$ times the penetration depth of an ordinary beam \footnote{The penetration depth of the ordinary beam is given by $\delta=c/(2n_i\omega)$} under the same conditions. The Fig. (3a) shows the component of the transmitted electric field, $E_{\phi}^T$, in 3D  intensity and we can note that the resulting wave keeps the desired characteristics when compared with $|F(z)|^2$ in Fig. (3b) and the Fig. (3b) shows the compensated incident beam. Here, we use $m=921$ and $g=341$ terms to get the aforementioned FWs.

%\h In Fig. (4d) we can see the lateral tracks which taper at the interface at $z=0$ and form the beam incident  that after the multiple reflections on the following interfaces will result in the beam presented at in Fig. (4a). The Fig. (4e) shows the same beam in logarithmic scale. The reflected intensity of ${E_\phi}$ in the first medium in 3D and in logarithmic scale is presented in Fig. (3e) and Fig. (3f).
%The transmitted and incident beam spectrum, $S(\beta_r)$ and $S(\beta_1)$, is shown in Fig. (5b) and Fig. (5e).

%\begin{figure}[h!]
 % \centering
  %\includegraphics[width=10cm]{Fig3.jpg}
%\caption{The amplitude spectra of the: a) transmitted beam and b) incident beam.}
%\end{figure}
%begin{document}
\begin{figure}[!htb]
	\centering
	\subfloat{
		\includegraphics[height=5cm]{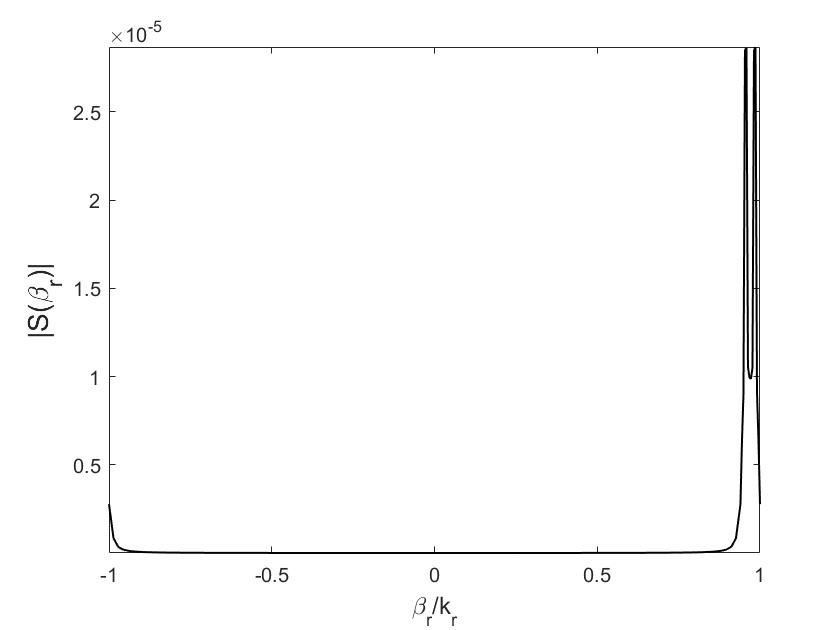}
 \label{(a)}}
	\quad %espaco separador
\subfloat{
\includegraphics[height=5cm]{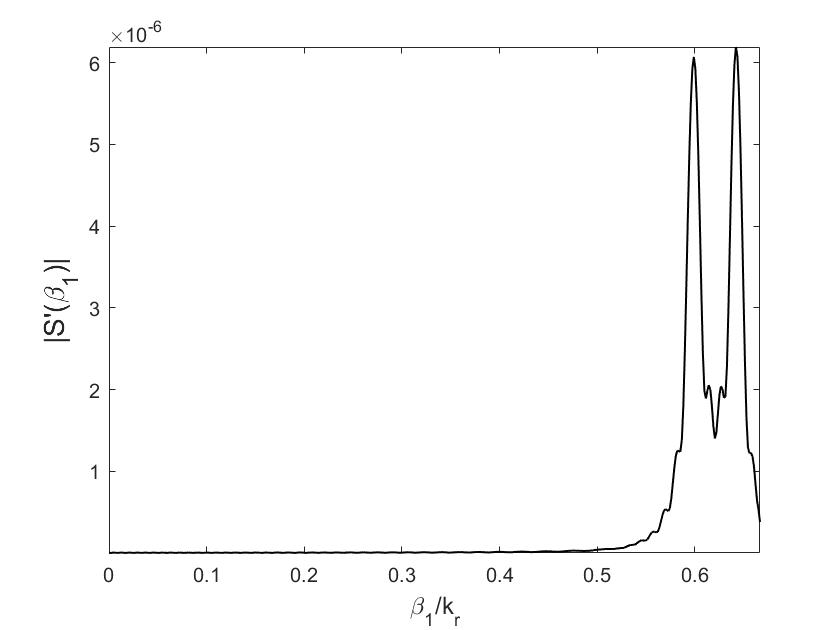}
 \label{(b)}}
\caption{The amplitude spectra of the: a) transmitted beam and b) incident beam.}
\end{figure}
 To conclude this section, we show in Figures (3a) and (3b) the amplitude spectra of the transmitted and incident beams, respectively.

\section{Conclusions}

 In this paper, we propose an analytical method for obtaining a highly non-paraxial beam with azimuthal polarization that, at normal incidence on an absorbing stratified medium, will provide in the last semi-infinite absorbing layer an azimuthally polarized and structured beam endowed with a micrometer intensity pattern chosen at will. The method also provides the beam incident on the first interface of the stratified medium that, ultimately, gives rise to the desired transmitted beam.

 We believe that the possibility of managing the properties of a highly non-paraxial beam under adverse conditions, such as multiple reflections in stratified structures and energy loss to the material media, may be of great importance in many different optical applications, like trapping and micro-manipulation, remote sensing, thin films, medical devices,medical therapies, etc.

\newpage
\large{\centering{\textbf{Appendix
\\
Effective transmission and reflection coefficients of a Bessel beam with azimuthal polarization in stratified media.}}}

%\chapter{\Large{\centering{Appendix}\\ \textbf{Effective transmission and reflection coefficients of a Bessel beam with azimuthal polarization in stratified media.}}}\\
%\\
%\textit{Grazielle de A. Lourenço-Vittorino and Michel Zamboni-Rached}

%\date{}
%\appendix
%\maketitle
 As we have shown in the paper, once the effective transmission coefficient of the azimuthally polarized beam incident upon the multilayered structure is known, we can obtain the desired structured beam in its last medium. Such coefficient, as well as the effective reflection coefficient, can be efficiently calculated through the transfer-matrix formulation, as we are going to see in this appendix.

%\h In reference [1], we present the matrix method to build scalar FWs from discrete Bessel beam superposition; here, we will adapt the results to a continuous superposition in view of the example discussed in the article  supplemented  by this appendix.
%\begin{figure}[!htb]
%	\centering
%	\subfloat{
%		\includegraphics[height=3cm]{apendice1.jpg}
%	}
	
%	\caption{Representation of a generic layer of a stratified media.}
%	\label{meio_estrat}
%\end{figure}

It is not difficult to show that the calculation of the effective transmission and reflection coefficients for an azimuthally polarized Bessel beam (hence, a TE beam) incident normally on a stratified medium, is similar to the calculation of the same coefficients in the case of a scalar plane wave also incident normally on the same medium and subject to the conditions that both it and its derivative must be continuous across each interface.
We will therefore reproduce here the obtaining of these coefficients in the case of a scalar plane wave.

Let us consider, in the stratified material, an arbitrary layer of thickness $L$ and, within this layer, a pair of plane waves\footnote{Here, for the sake of simplicity, we suppress the harmonic time variation term $\exp(-i\omega t)$} (propagating and counterpropagating):

%\h So, let us assume that the total field formed by an single Bessel beam with azimuthal polarization and longitudinal wavenumber $\beta$ in a generic layer of a stratified medium of thickness $L$ is given by:
%
%\begin{equation}
%    E_{\phi}=J_1(h\rho)g(z)\hat{\phi}
%    \label{ephi}
%\end{equation}
%with
%\begin{equation}
%h=\sqrt{n^2k_0^2-\beta^2}
%\end{equation}
%where $J_1(.)$ is the first-order Bessel function, $k_0=\omega/c$ and $n$ is the refractive index of the medium. Furthermore,

\begin{equation}
g(z)=B^r\exp(i\beta z)+B^l\exp(-i\beta z)
\label{gzinho}
\end{equation}
whose derivative with respect to $z$ is
\begin{equation}
G(z)=i\beta B^r\exp(i\beta z)-i\beta B^l\exp(-i\beta z)
\label{gzao}
\end{equation}
where $\beta = n\omega/c $, being $n$ the refractive index of the layer, $B^r$ and $B^l$ are constants with the superscripts $r$ and $l$ indicating the beam propagating to the right and the left directions, respectively.

The boundary conditions are the continuity of $g$ and its derivative $G$ through any interface. That means that by knowing their values on an interface at $z$, we can get them on the next interface at $z+L$. So, by using $(\ref{gzinho})$ and $(\ref{gzao})$, we write:

%\h Given that the transverse wavenumber $h$ of the forward and backward Bessel beams is conserved, it can be noted that the continuity of the tangential component of the electric field $E_{\phi}$ (and obviously of the tangential component of field $H_{\phi}$ which was not addressed here but can be available through Faraday's law) is conditioned by $g(z)$ and $G(z)$. This means that knowing $g(z)$ and $G(z)$ on the interface at  $z=z$, we can get its on the next interface $z=z+L$. Thereby, using $(\ref{gzinho})$ and $(\ref{gzao})$, we write:

\begin{equation}
g(z+L)=B^r\exp(i\beta L)\exp(i\beta z)+B^l\exp(-i\beta L)\exp(-i\beta z)
\label{gzinho_zl}
\end{equation}

\begin{equation}
G(z+L)=i\beta B^r\exp(i\beta l)\exp(i\beta z)-i\beta B^l\exp(-i\beta L)\exp(-i\beta z)
\label{gzao_zl}
\end{equation}

Those fields at $z$ and $z+L$ can be related by a matrix in the following way

\begin{equation}
 \left[ \begin{array}{cc}
	g(z+L) \\
	G(z+L) \end{array} \right] \ug
=\texttt{\textbf{M}}\,\left[
\begin{array}{cc}
	g(z) \\
	G(z)\end{array} \right] \,\,,
\label{M1}
\end{equation}
with
\begin{equation}  \texttt{\textbf{M}}= \left[ \begin{array}{cc}
	\texttt{\textbf{M}}_{11} & \texttt{\textbf{M}}_{12} \\
	\texttt{\textbf{M}}_{21} & \texttt{\textbf{M}}_{22}
\end{array} \right] \,\, . \label{M2} \end{equation}

Substituting ($\ref{gzinho}$)-($\ref{gzao_zl}$) in ($\ref{M1}$), we obtain the unimodular matrix:

\begin{equation}
\texttt{\textbf{M}}\ug =\left[ \begin{array}{cc}
	\cos(\beta L) & \frac{1}{\beta}\sin(\beta L) \\
	-\beta\sin(\beta L) &\cos(\beta L) \,\, .
\end{array} \right] \label{matrix}
\end{equation}

If the structure has $M$ layers, i.e., $m=1,2,..,M$, then $\beta\rightarrow \beta_m$ and
  $L\rightarrow L_m$, such that

  \begin{equation}
\texttt{\textbf{M}}_m\ug =\left[ \begin{array}{cc}
	\cos(\beta_{m}L_m) & \frac{1}{\beta_{m }}\sin(\beta_{m}L_m) \\
	-\beta_{m}\sin(\beta_{m }L_m) &\cos(\beta_{m }L_m)
\end{array} \right] \label{matrix}
\end{equation}
and finally the resulting matrix will be
\begin{equation}  \texttt{\textbf{M}}=\texttt{\textbf{M}}_{M-1}\cdot...\cdot
\texttt{\textbf{M}}_3\cdot \texttt{\textbf{M}}_2  = \left[ \begin{array}{cc}
	\texttt{\textbf{M}}_{11} & \texttt{\textbf{M}}_{12} \\
	\texttt{\textbf{M}}_{21} & \texttt{\textbf{M}}_{22}
\end{array} \right] \,\, , \label{M} \end{equation}
\begin{figure}[!htb]
	\centering
	\subfloat{
		\includegraphics[height=3cm]{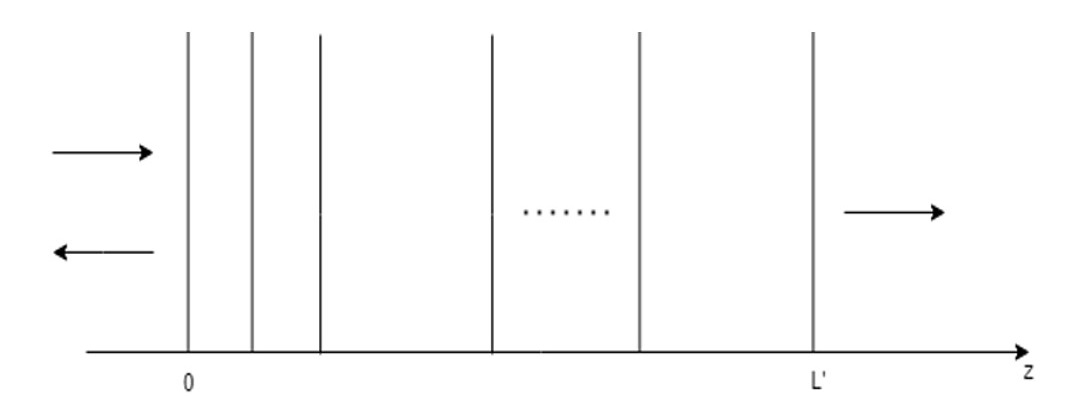}
	}
	
	\caption{Representation of a stratified media.}
	\label{meio_estrat}
\end{figure}

%\h Now, to obtain the effective  transmission $\tau$ and reflection $\gamma$ coefficients, let us assume the scenario illustrated in Fig.(1), where the plane wave coming from the first medium impinges normally upon the first interface, undergoes multiple reflections and transmissions  in the following interfaces and a fraction of it is transmitted to the last medium, maintaining the same order and transverse wavenumber of the incident beam. Thus, the fields in the semi-infinite media to the right and to the left side of the stratified media are given as shown bottom:

Now, to obtain the effective  transmission $\tau$ and reflection $\gamma$ coefficients, let us assume the scenario illustrated in Fig.(1), where a plane wave coming from the first medium impinges normally upon the first interface, undergoes multiple reflections and transmissions  in the following interfaces and a fraction of it is transmitted to the last medium. Thus, the fields in the semi-infinite media to the right and to the left side of the stratified structure are given by:

\begin{equation}
g_l(z)=B_0^r\exp(i\beta_l z)+\gamma B_0^r\exp(-i\beta_l z)
\label{g_2}
\end{equation}

\begin{equation}
G_l(z)=i\beta_l B_0^r\exp(i\beta_l z)-i\beta_l \gamma B_0^r\exp(-i\beta_l z)
\label{G_2}
\end{equation}

\begin{equation}
g_r(z)=\tau B_0^r\exp(i\beta_r z)
\label{g_3}
\end{equation}

\begin{equation}
G_r(z)=i\beta_r \tau B_0^r\exp(i\beta_r z)
\label{G_3}
\end{equation}

 Now, let us assume that the structure starts at $z=0$ and ends at $z=L'$. So, from the $(\ref{g_2})$-$(\ref{G_3})$ we have:

\begin{equation}
g_l(0)=B_0^r+\gamma B_0^r
\label{gzinho_0}
\end{equation}

\begin{equation}
G_l(0)=i\beta_l B_0^r-i\beta_l \gamma B_0^r
\label{gzao_0}
\end{equation}
and

\begin{equation}
g_r(L')=\tau B_0^r\exp(i\beta_r L')
\label{gzinho_l}
\end{equation}

\begin{equation}
G_r(L')=i\beta_r \tau B_0^r\exp(i\beta_r L')
\label{gzao_l}
\end{equation}

 By using the transfer matrix:
\begin{equation}
 \left[ \begin{array}{cc}
	g_r(L') \\
	G_r(L') \end{array} \right] \ug
=\texttt{\textbf{M}}\,\left[
\begin{array}{cc}
	g_l(0) \\
	G_l(0)\end{array} \right] \,\,,
\label{eqmatrix2}
\end{equation}
and replacing Eqs.$(\ref{gzinho_0})$-$(\ref{gzao_l})$ in Eq.$(\ref{eqmatrix2})$, we then obtain $\tau$ e $\gamma$:

\begin{equation}
\tau=2i\beta_{1}e^{-i\beta_{M}L'}\left[\frac{1}{-\texttt{\textbf{M}}_{21}+\beta_{1}\beta_{M}
	\texttt{\textbf{M}}_{12}+i(\beta_{1}\texttt{\textbf{M}}_{22}+\beta_{M}\texttt{\textbf{M}}_{11})}\right] \label{tauM}
\end{equation}

\begin{equation}
\gamma=\frac{(\texttt{\textbf{M}}_{21}+\beta_{1}\beta_{M}\texttt{\textbf{M}}_{12})+i(\beta_{1}\texttt{\textbf{M}}_{22}-\beta_{M}\texttt{\textbf{M}}_{11})}{(-\texttt{\textbf{M}}_{21}+\beta_{1}\beta_{M}\texttt{\textbf{M}}_{12})+i(\beta_{1}\texttt{\textbf{M}}_{22}+\beta_{M}\texttt{\textbf{M}}_{11})} \label{gama1q}
\end{equation}

Finally, it is important to say that in the case of an azimuthally polarized Bessel beam, as addressed in the paper, the equations for the effective reflection and transmission coefficients are still given by Eqs.(\ref{tauM},\ref{gama1q},\ref{matrix},\ref{M}), but in them $\beta_m = \sqrt{(n_m\omega/c)^2 - h^2}$, where $h$ is the transverse wave number of the incident Bessel beam and which is conserved throughout the stratified structure.

%\begin{comment}
%\begin{thebibliography}{99}

%	\bibitem{Lourenco-Vittorino} G De A. Lourenço-Vittorino and M.Zamboni-Rached:``Modeling the longitudinal intensity pattern of diffraction resistant beams in stratified media", Applied Optics 57 (20) 5643 (2018) .
%\end{thebibliography}

\end{document}